\documentclass[nofootinbib, 10pt]{revtex4-2}
\usepackage{graphicx}
\usepackage[T1, T2A]{fontenc}
\usepackage[utf8]{inputenc}
\usepackage[english]{babel}
\usepackage{amsmath}
\usepackage{amssymb}
\usepackage{perpage}
\MakePerPage{footnote}
\usepackage{caption}
\usepackage{comment}
\usepackage[lofdepth, lotdepth]{subfig}
\usepackage{float}
\usepackage{multirow}
\usepackage{stackengine}[2013-10-15]
\usepackage{dutchcal}
\usepackage[normalem]{ulem}
\usepackage{xcolor}

\addto\captionsenglish{}
\begin{document}

\preprint{INR-TH-2025-021}

\title{Towards spinning $U(1)$ gauged non-topological solitons in the model with Chern--Simons term}

\author{Ivan Ivashkin}
  \email{ivashkin.ir@phystech.edu}
  \affiliation{Moscow Institute of Physics and Technology, Institutsky lane 9, Dolgoprudny, Moscow region, 141700}
  \affiliation{Institute for Nuclear Research of RAS, prospekt 60-letiya Oktyabrya 7a, Moscow, 117312}

\author{Eduard Kim}
  \affiliation{Moscow Institute of Physics and Technology, Institutsky lane 9, Dolgoprudny, Moscow region, 141700}
  \affiliation{Institute for Nuclear Research of RAS, prospekt 60-letiya Oktyabrya 7a, Moscow, 117312}

\author{Emin Nugaev}
  \affiliation{Institute for Nuclear Research of RAS, prospekt 60-letiya Oktyabrya 7a, Moscow, 117312}

\author{Yakov Shnir}
  \affiliation{BLTP JINR, Joliot--Curie St 6, Dubna, Moscow region, 141980}

\begin{abstract}
    We obtain localized field configurations with finite energy in a ($2+1$)-dimensional model with Maxwell and Chern--Simons gauge terms coupled to a massive complex scalar field. These non-topological solitons are characterized by the $U(1)$ frequency and a winding number. Thus, the solutions possess Noether charge and non-trivial angular momentum, which is not quantized in contrast to the topological case. We study the solitons and their integral characteristics numerically and demonstrate that they are kinematically stable. The obtained solutions allow for the thin-wall approximation in some region of frequencies. For each winding number, the Noether charge has a lower bound that coincides with an isolated point, where the non-relativistic conformal symmetry seems to be restored.
\end{abstract}

\maketitle


\section{Introduction}
The study of various solitonic configurations in gauge field theories has a long and rich history, providing deep insight both into non‑perturbative phenomena and into the interplay between topology, gauge structure, and matter fields; see, e.g. \cite{manton2004topological,Shnir:2018yzp}. In particular, models in two spatial dimensions with Chern--Simons interactions have attracted sustained interest \cite{Paul:1986ix,Polyakov:1988md,Jackiw:1990tz,Dunne:1990pp,Ghosh:1993yz,Arthur:1996uu,Torres:1992wz,Navarro-Lerida:2016omj,Blazquez-Salcedo:2013roa,Navarro-Lerida:2008dcl,Loginov:2014ora,Loginov:2018wbo}.

The presence of a Chern--Simons term in the gauge sector alters the dynamics and can lead to unusual features such as fractional spin and anyonic statistics \cite{Polyakov:1988md,Forte:1990hd}, generation of topological mass of the gauge field \cite{Deser:1981wh} and allowing for existence of electrically charged topologically nontrivial vortices \cite{Paul:1986ix,Inozemtsev:1987wv,Zhang:1992eu,Horvathy_2009,Bazeia:2012ux,Bazeia:2017vzq,Andrade:2020sbl}. The Chern--Simons theory provides a topological explanation for the quantization of the Hall conductivity \cite{Ishikawa:1983ad}.

Apart from Chern--Simons vortices in the generalized Abelian Higgs model, there is another interesting example of topological solitons, the planar skyrmions \cite{Bogolubskaya:1989ha,Bogolyubskaya:1989fz,Leese:1989gi,Piette:1994jt}, which may provide a specific contribution in the description of the topological quantum Hall effect \cite{PhysRevLett.102.186602}. Vortices in the Maxwell--Chern--Simons--Skyrme model were investigated in \cite{Loginov:2014ora,Samoilenka:2016wys,Navarro-Lerida:2023fsr}.

In the present work, we explore a class of stationary spinning tube‐shaped localized field configurations in a ($2+1$)-dimensional model which combines both the Maxwell and the Chern--Simons gauge terms, coupled to a complex scalar field. Non-topological solitons in this model carry electric charge and magnetic flux, and they also possess an angular momentum.

The resulting system brings together several interesting aspects. The potential of the scalar field supports the usual ungauged ($2+1$)-dimensional Q-ball solutions \cite{Volkov:2002aj} in the decoupled limit. However, presence of the Chern--Simons term alters the familiar energy-charge and charge-momentum relations compared to ordinary solutions in the same way, as it happens in the Maxwell--Chern--Simons--Skyrme model \cite{Loginov:2014ora,Navarro-Lerida:2016omj,Samoilenka:2016wys}.

We start with introducing the model and analytical consideration of spinning non-topological solitons in Sec.~\ref{Model}. Numerical analysis of the solitonic solutions and their properties is provided in Sec.~\ref{Numerical}. We discuss our results in Sec.~\ref{Outlook}.

\section{The model}\label{Model}

Let us consider a ($2+1$)-dimensional Lorentz-invariant theory of the massive complex scalar field $\varphi$ and the Abelian gauge field $A^\mu$. The Lagrangian of the theory is\footnote{We use the natural system of units: $\hbar=c=1$. The Minkowski metric in ($2+1$)-dimensional space-time is mostly negative: $\eta_{\mu\nu}=\text{diag}(1,-1,-1)$.} \cite{Deshaies-Jacques:2006clf}
\begin{equation}\label{Lagr}
    \mathcal{L}=-\frac{1}{4}F_{\mu\nu}F^{\mu\nu} +\frac{g}{2} \epsilon^{\mu\nu\rho}A_\mu\partial_\nu A_\rho+D_\mu\varphi D^\mu\varphi^*-m^2\varphi\varphi^*+\frac{\lambda}{2}(\varphi\varphi^*)^2-\frac{\sigma}{3}(\varphi\varphi^*)^3,
\end{equation}
where $F_{\mu\nu}=\partial_\mu A_\nu-\partial_\nu A_\mu$ and the covariant derivative is written as $D_\mu=\partial_\mu-ieA_\mu$. The parameters of the theory have the following mass dimensions: $[m]=[g]=[\lambda]=1$, $[\sigma]=0$, $[e]=1/2$; furthermore, both $\lambda$ and $\sigma$ are positive-defined. Unlike scalar electrodynamics, the Lagrangian includes a Chern--Simons term $\epsilon^{\mu\nu\rho}A_\mu\partial_\nu A_\rho$ ($\epsilon^{012}=1$), which is gauge-invariant. Therefore, the theory~(\ref{Lagr}) possesses a local $U(1)$ symmetry. Note that the Chern--Simons term breaks the P symmetry in ($2+1$)-dimensional space-time.

To obtain spinning non-topological solitons, we introduce the following ansatz\footnote{In contrast to Greek indices, which are Minkowski space-time indices, Latin indices are spatial indices, which are summed over using Euclidean signature, and $A^\mu=\big(A^{(0)},A^i\big)$. In addition, we set $\epsilon^{12}=1$.}:
\begin{equation}\label{Ansatz}
    \varphi(t,\vec{x})=e^{-i\omega t+in\theta}f(r),\qquad A^{(0)}(t,\vec{x})=G(r),\qquad
    A^i(t,\vec{x})=-\epsilon^{ij}\frac{a(r)x^j}{r^2}.
\end{equation}
Here $r=\sqrt{\vec{x}^2}$ and $\theta$ are polar coordinates; $f=f(r)$, $G=G(r)$, $a=a(r)$ are axially symmetric smooth real-valued functions; $\omega$ is a parameter of the solution interpreted as the $U(1)$ frequency, and $n\in\mathbb{Z}$ is a winding number. Note that the anzatz (\ref{Ansatz}) describes a massive complex scalar field with a harmonic dependence on time $t$ and on the angular coordinate $\theta$. For $n=0$ the system (\ref{Lagr}) is reduced to the case considered in \cite{Deshaies-Jacques:2006clf}. 

Corresponding equations of motion are (prime means the derivative with respect to $r$):
\begin{gather}
    f''+\frac{f'}{r}+\bigg(\omega^2-m^2-\frac{n^2}{r^2}\bigg)f+\lambda f^3-\sigma f^5=e^2f\bigg(\frac{a^2}{r^2}-G^2\bigg)-2ef\bigg(\frac{na}{r^2}+\omega G\bigg),\label{f_eq}\\[0.2cm]
    G''+\frac{G'}{r}=2ef^2(\omega +eG)-g\frac{a'}{r},\qquad a''-\frac{a'}{r}=2ef^2(ea-n)-g rG'.\label{Ga_eq}
\end{gather}
Note that Eqs.~(\ref{f_eq}, \ref{Ga_eq}) have a discrete symmetry: they are invariant with respect to $g\to -g$, $n\to -n$, $a\to -a$ transformations. One can specify another similar discrete symmetry: $g\to -g$, $\omega\to -\omega$, $G\to -G$. Without loss of generality, we assume $\omega >0$.

The electric and magnetic components of the field tensor $F_{\mu\nu}$ are expressed by the following formulas:
\begin{equation}\label{EH_eq}
    \mathcal{E}^i(r)=-G'(r)\frac{x^i}{r},\qquad \mathcal{H}(r)=\epsilon_{ij}\partial_iA_j=\frac{a'(r)}{r}.
\end{equation}
In this notation, the energy of the field configuration is
\begin{equation}\label{Energy}
    E=2\pi\int\limits_{0}^\infty\Bigg[\frac{\vec{\mathcal{E}}^2+\mathcal{H}^2}{2}+(f')^2+\bigg(\omega^2+m^2+\frac{n^2}{r^2}\bigg)f^2-\frac{\lambda}{2}f^4+\frac{\sigma}{3}f^6+eGf^2(eG+2\omega)+e\frac{af^2}{r^2}(ea-2n)\Bigg]r\mathop{dr}.
\end{equation}
In accordance with \cite{Navarro-Lerida:2016omj}, the integrand from Eq.~(\ref{Energy}) does not include any contribution coming from the Chern--Simons term $\epsilon^{\mu\nu\rho}A_\mu\partial_\nu A_\rho$. This can be explained by the topological nature of the corresponding term in the action $\mathcal{S}=\int \mathcal{L}\sqrt{|g|}\mathop{d^3x}$ considered when varying with respect to the metric $g_{\mu\nu}$. The energy functional (\ref{Energy}) is the same as in the case of scalar electrodynamics, as expected. Note that the electric field energy $\int\vec{\mathcal{E}}^2/2\cdot r\mathop{dr}$ diverges in pure Maxwell theory for a point-like charge at infinity, and this is the crucial obstacle for the existence of solitons. In our model, the divergence is cured by the Chern--Simons term.

The requirements of regularity and finiteness of the energy lead to the following boundary conditions for the function~$f$:
\begin{align}
    f'\big|_{r=0}=0,\qquad f\big|_{r=\infty}=0&\qquad\text{in case of}\quad n=0;\label{bc_f0}\\[0.15cm]
    f\big|_{r=0}=0,\qquad f\big|_{r=\infty}=0&\qquad\text{in case of}\quad n\neq 0.\label{bc_fneq0}
\end{align}
For the functions $G$ and $a$ we have (for all $n$):
\begin{align}
    \label{bc_G}
    G'\big|_{r=0}=0,\qquad & G\big|_{r=\infty}=0,\\[0.15cm]
    \label{bc_a}
    a\big|_{r=0}=0,\qquad & a'\big|_{r=\infty}=0.
\end{align}
The equations of motion (\ref{f_eq}, \ref{Ga_eq}) supplemented with the boundary conditions above cannot be solved analytically, and their numerical solution will be discussed in the next section. Nevertheless, one can show that the functions $f$, $G$, and $a$ approach their asymptotic values at $r\to\infty$ exponentially (see App.~\ref{app_asymp}). In other words, presence of the Chern--Simons term affects the equations of motion (\ref{f_eq}, \ref{Ga_eq}), so that all the fields, including $\vec{\mathcal{E}}$, decrease exponentially at spatial infinity. Thus, regularization of the energy (\ref{Energy}) is achieved: $\int\vec{\mathcal{E}}^2/2\cdot r\mathop{dr}<\infty$.

It is well known that in scalar field theories, stabilization of non-topological solitons is ensured by the conservation of the global $U(1)$ charge \cite{Shnir:2018yzp}. In the present case, the charge has the following form in terms of ansatz functions:
\begin{equation}\label{Charge}
    Q=2\pi \int\limits_{0}^\infty\big[2f^2(\omega+eG)\big]r\mathop{dr}. 
\end{equation}
Using Eqs.~(\ref{Energy}) and (\ref{Charge}), one can obtain a well-known relation
\begin{equation}\label{EQ_relation}
    \frac{\partial E}{\partial\omega}=\omega\frac{\partial Q}{\partial\omega},\qquad\text{or}\qquad\frac{\partial E}{\partial Q}=\omega,
\end{equation}
which is valid for non-topological solitons in many other models \cite{Nugaev:2019vru}. We use it to check the correctness of our numerical calculations.

In addition to the charge (\ref{Charge}), the solution under consideration carries the magnetic flux
\begin{equation}\label{mag_flux}
    \Phi=\int\limits_{\mathbb{R}^2}\mathcal{H}\mathop{d^2x}=2\pi \int\limits_{0}^\infty a'\mathop{dr}=2\pi\cdot a\big|_{r=\infty},
\end{equation}
that is also conserved. In contrast to the case of topological vortices \cite{Horvathy_2009}, the magnetic flux is not quantized. However, the relation between $U(1)$ charge and magnetic flux, which is specific to vortex solutions, remains valid:
\begin{equation}\label{Q_Phi}
    Q=\frac{g}{e}\Phi.
\end{equation}
This formula is obtained by integrating the first of Eqs.~(\ref{Ga_eq}) in light of the boundary conditions (\ref{bc_G}, \ref{bc_a}).

Presence of the gauge field, as well as angular dependence of the ansatz for the scalar field, provides nontrivial angular momentum $J$:
\begin{equation*}
    J=\int\limits_{\mathbb{R}^2}\epsilon^{ij}x^iT^{0j}\mathop{d^2x}=-2\pi\int\limits_0^\infty\big[2f^2(\omega+eG)(n-ea)-G'a'\big]r\mathop{dr},
\end{equation*}
where $T^{\mu\nu}$ is a symmetric energy-momentum tensor. Using Eq.~(\ref{Charge}), the first of Eqs.~(\ref{Ga_eq}), and the relation (\ref{Q_Phi}), the last expression can be reduced to the following form after integrating by parts:
\begin{equation}\label{Ang_momentum}
    J=-nQ+\frac{e^2}{4\pi g}Q^2.
\end{equation}
Note that the angular momentum has no quantization condition and its sign is not fixed: one can easily change it by replacing $g\to -g$, $n\to -n$, $a\to -a$ (as already mentioned, these transformations do not change the equations of motion). Furthermore, the function $J=J(Q)$ has a global extremum at $Q=2\pi gn/e^2$, which is a minimum in the current notation. This is consistent with the results for rotating gauged skyrmions reported in \cite{Loginov:2014ora}.

Thus, the solitonic solution possesses both the Noether charge $Q$ and angular momentum $J$ for any winding numbers $n$ in the model with Chern--Simons term (\ref{Lagr}).

\section{Numerical solutions and their properties}\label{Numerical}

In this section, we present numerical solutions of Eqs.~(\ref{f_eq}, \ref{Ga_eq}) and some of their properties. First of all, we rescale all the fields and parameters using the scalar field mass $m$ as an energy unit (e.g. $\tilde f=f/m^{1/2}$, etc.). The dimensionless equations of motion can be obtained from Eqs.~(\ref{f_eq}, \ref{Ga_eq}) by setting $m=1$. Therefore, we have a boundary value problem comprising Eqs.~(\ref{f_eq},~\ref{Ga_eq}) and the boundary conditions (\ref{bc_f0}--\ref{bc_a}).

\begin{figure}[H]
	\includegraphics[width=1.0\linewidth]{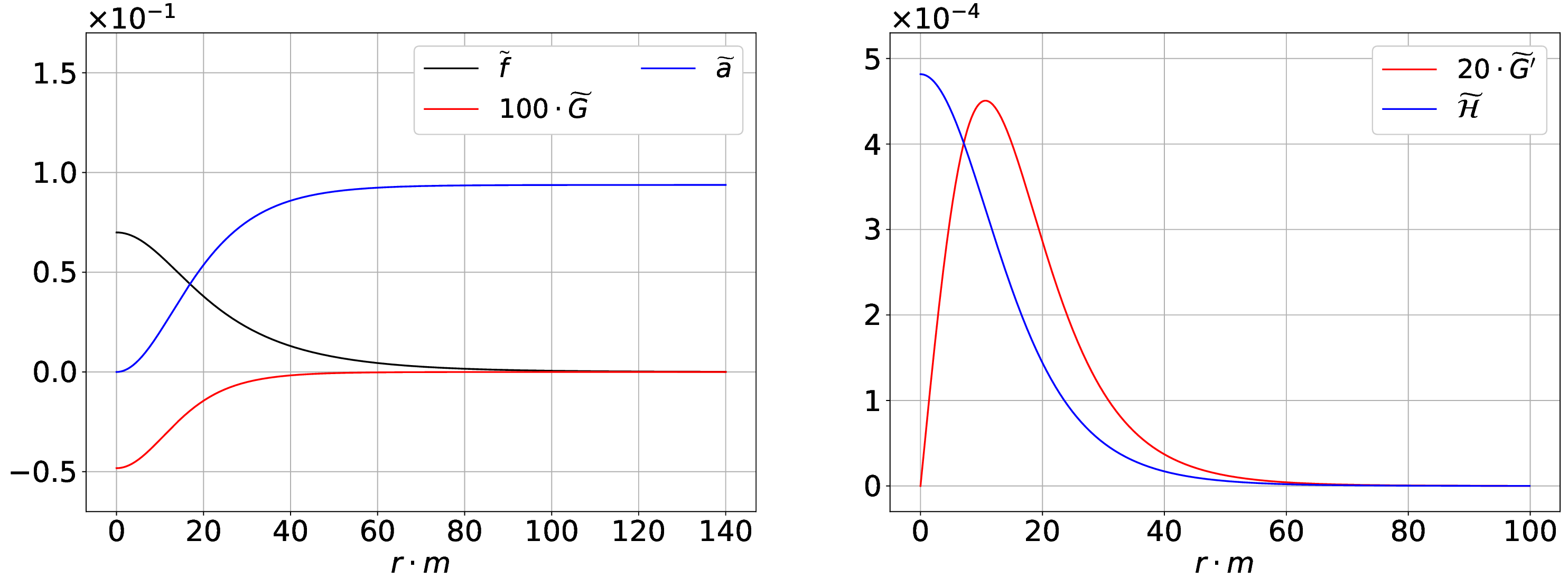}
    \caption{Profiles of the soliton with $n=0$. \textit{Left~panel:} Amplitudes of the fields $\varphi$, $A^\mu$. \textit{Right~panel:} Amplitudes of the fields $\vec{\mathcal{E}}$, $\mathcal{H}$. Here $g/m=1$, $\lambda/m=2$, $\sigma=3/2$, $e/\sqrt{m}=0.05$, and $\omega/m=0.999$.}\label{fig_n=0}
\end{figure}

\begin{figure}[H]
	\includegraphics[width=1.0\linewidth]{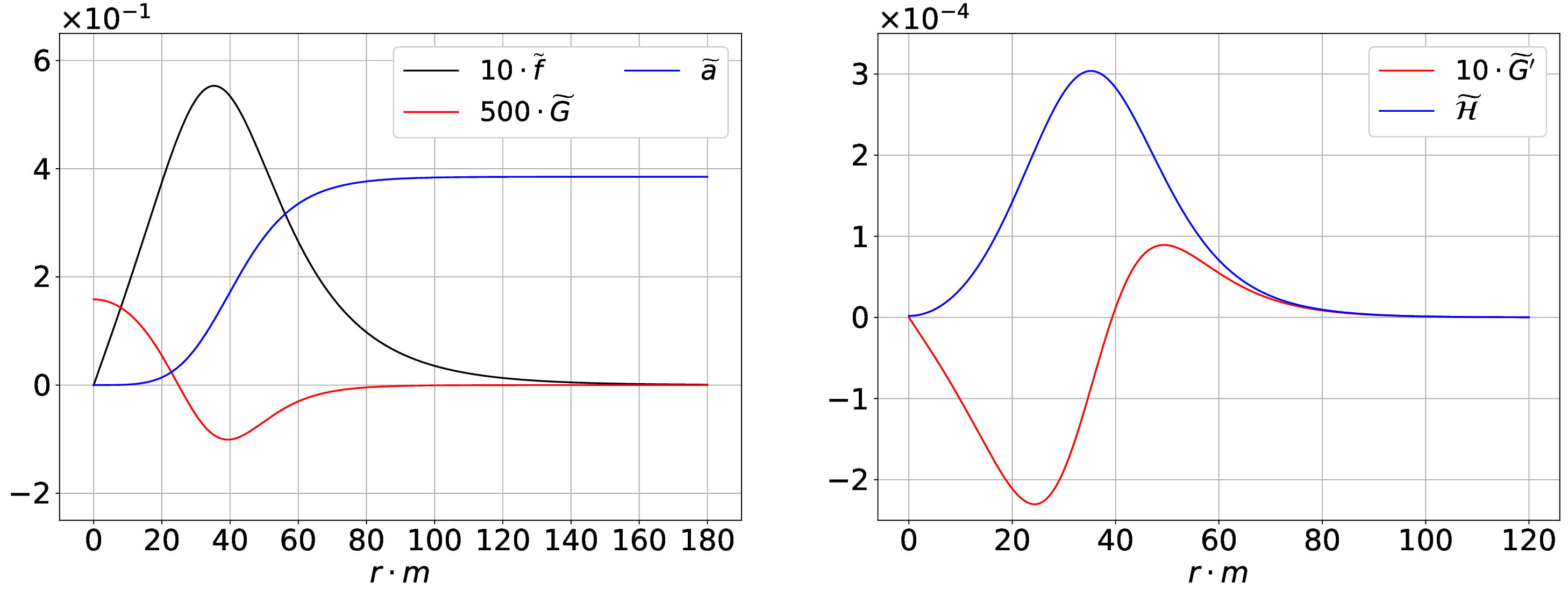}
    \caption{Profiles of the soliton with $n=1$. \textit{Left~panel:} Amplitudes of the fields $\varphi$, $A^\mu$. \textit{Right~panel:} Amplitudes of the fields $\vec{\mathcal{E}}$, $\mathcal{H}$. Here $g/m=1$, $\lambda/m=2$, $\sigma=3/2$, $e/\sqrt{m}=0.05$, and $\omega/m=0.999$.}\label{fig_n=1}
\end{figure}

We solve the boundary value problem numerically for different winding numbers $n$ using a fourth-order collocation method\footnote{The method is based on the algorithm described in \cite{Kierzenka:2001} and has a SciPy \cite{Virtanen:2019joe} implementation.}. To check the correctness of our numerical results, we repeat the calculations by employing a professional solver \cite{schoen}, with typical errors of the order of $10^{-5}$. It also allows us to check that rotational invariance of the solutions persists for all range of values of $n$. The calculations are performed at the following parameters of the model: $\tilde\omega=\omega/m=0.999$, $\tilde\lambda=\lambda/m=2$, $\tilde\sigma=\sigma=3/2$, $\tilde e=e/\sqrt{m}=0.05$, and $\tilde g=g/m=1$. In addition, we use formulas (\ref{EH_eq}) to find the amplitudes of the electric and magnetic fields. The obtained solutions are presented in Figs.~\ref{fig_n=0}, \ref{fig_n=1} (for $n=0$ and $n=1$, respectively). 

As we see, the field configurations are localized. Their energy and charge can be calculated by the formulas (\ref{Energy},~\ref{Charge}). The localization is provided by the Chern--Simons term since the coupling constant $g$ plays the role of the gauge field mass, causing the amplitudes $G$ and $a$ to decay exponentially at spatial infinity.

\begin{figure}[H]
    \includegraphics[width=1.0\linewidth]{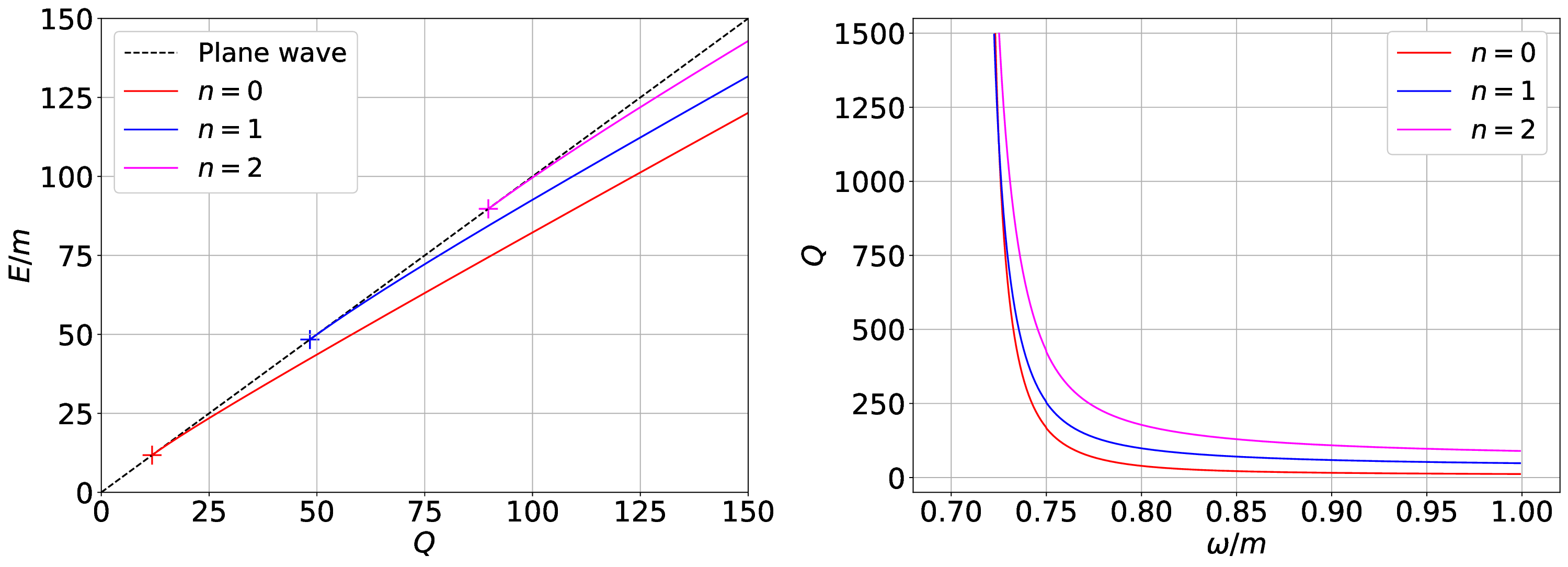}
    \caption{\textit{Left~panel:} Soliton energy $E/m$ as a function of its $U(1)$ charge $Q$ for different winding numbers $n$. \textit{Right~panel:}~Noether charge $Q$ as a function of frequency $\omega/m$ for the solitonic solutions with different winding numbers $n$. Here $g/m=1$, $\lambda/m=2$, $\sigma=3/2$, and $e/\sqrt{m}=0.05$.}
    \label{E(Q)}
\end{figure}

By numerical differentiation with respect to the frequency $\omega$, it is possible to verify that obtained solutions satisfy the relation (\ref{EQ_relation}). Consequently, the parameter $\omega$ can be considered as a chemical potential for a system of the particles. As far as $\omega>0$, $\partial E/\partial Q>0$. Thus, the function $E=E(Q)$ is growing, as can also be seen from the plot (see left panel of Fig.~\ref{E(Q)}).

In addition, the $E=E(Q)$ plot shows that considering field configurations are kinematically stable since $E<mQ$. It means that the energy of the system is less than the energy of free particles with the same charge. Therefore, the solitonic solution corresponds to a bound state of the system, and the quantum-mechanical fission into scalar particles is energetically forbidden.

From the left panel of Fig.~\ref{E(Q)} it can also be seen that $\partial^2E/\partial Q^2<0$, which is equivalent to the relation
\begin{equation}\label{Vakh_Kol}
    \frac{\partial Q}{\partial\omega}=\frac{1}{\omega}\frac{\partial E}{\partial\omega}<0.
\end{equation}
In the context of models without gauge field, the last inequality is commonly referred to as a Vakhitov--Kolokolov stability criterion \cite{Vakhitov:1973}. However, presence of the gauge field complicates matters \cite{Panin:2016ooo} and does not allow us to assert that the relation (\ref{Vakh_Kol}) indicates the linear stability of the solitons.

Our numerical procedure includes iterative decreasing of the frequency $\omega$. The results of the calculations indicate existence of a minimum value $\omega_\text{min}$, such that a limit $\omega\to\omega_\text{min}$ corresponds to a thin-wall regime \cite{Shnir:2018yzp}. One can easily estimate the value of $\omega_\text{min}$ analytically, supposing the smallness of the coupling constant $e$: $e/\sqrt{m}\ll 1$. In this case, the gauge field has a small effect on $\omega_\text{min}$, and it is possible to use a relation
\begin{equation*}
    \omega_\text{min}^{(\text{sc})}=\sqrt{m^2-\frac{3\lambda^2}{16\sigma}},
\end{equation*}
which is valid for the pure scalar model \cite{Coleman:1985ki}. The relation leads to the following estimation (at the selected $m$, $\lambda$, and~$\sigma$): $\omega_\text{min}\simeq \omega_\text{min}^{(\text{sc})}=1/\sqrt{2}$. This is consistent with our numerical results.

At the limit $\omega\to\omega_\text{min}$, the soliton grows; thus, the energy $E$ and charge $Q$ take large values (see right panel of Fig.~\ref{E(Q)}). Note that the thin-wall approximation becomes applicable at the charges of the order of $7,5\cdot 10^2$ (for $n=1$). Considering the behavior of the function $E=E(Q)$, one can numerically check that $E$ is proportional to $Q$: $E\propto \omega Q$ ($\omega\to\omega_\text{min}$). Note that for the entire set of solutions, the charge $Q$ can be a multi-valued function of the frequency $\omega$, as it was pointed out for $n=0$ in \cite{Deshaies-Jacques:2006clf}. In this study, we restrict our consideration to the first branch of solutions with nontrivial winding numbers $n$, since  analysis of the properties of solutions on the second, higher in charge, branch requires application of other numerical methods.

One of the remarkable properties of the studied solutions is associated with angular momentum $J$. According to Eq.~(\ref{Ang_momentum}), its minimum value $J_n=-\pi n^2g/e^2\equiv-nQ_n/2$ is achieved at $Q=Q_n\equiv 2\pi gn/e^2=g\,\Phi_n/e$, where $\Phi_n=2\pi n/e\equiv n\Phi_0$. It is known that such a relation between the charge $Q_n$ and the quantized magnetic flux $\Phi_n$ is typical for vortices \cite{Horvathy_2009}. In the present case, the relation is true for the solution carrying an extremum value of the angular momentum, but, in general, the charge can take arbitrary values. It is also noteworthy that there is a value of the charge for which the angular momentum vanishes: $J=0$, if $Q= 4\pi g n/e^2=2Q_n$. Note that $Q_n\simeq 2.51\cdot 10^3\cdot n$ at the selected $e$ and $g$. In particular, in our model $1\ll Q_1<Q\big|_{\omega\,=\,\omega_\text{min}}$, and the minimum value of $J$ is achievable, as evidenced by the numerically check for $n=1$.

On the other hand, some significant features of the solitonic solutions can be noted in the region of relatively small charges, considering in the $\omega\to m$ limit. In this case, a quadratic term from Eq.~(\ref{Ang_momentum}) acts as a small correction since $e^2/(4\pi g)\simeq 10^{-4}$ and $Q\lesssim 10^2$ (at the selected parameters); thus, $J\simeq -nQ$, that reproduces a similar relation for spinning Q-balls \cite{Volkov:2002aj}.

Moreover, at $\omega=m$, a kinematically stable branch of the non-topological solitons merges the plane-wave line $E=mQ$, ending at an isolated point (see left panel of Fig.~\ref{E(Q)}; isolated points are marked with crosses). Such a type of behavior is typical for non-relativistic models with conformal invariance \cite{Galushkina:2025hkw, Galushkina:2025yce}. Therefore, studying the limit $\omega\to m$ in case of gauged theory is of interest due to the richer symmetry group.

\section{Outlook}\label{Outlook}

In this paper, we have studied the $(2+1)$-dimensional Maxwell--Chern--Simons model. In the model, a gauge field is coupled to a massive complex scalar field, carrying a $U(1)$ charge. Using a spinning ansatz, that includes a $U(1)$ frequency $\omega$ and a winding number $n$ as parameters, non-topologial solitons were obtained. For any winding numbers, the solitonic solutions possess the Noether charge $Q$ and the non-trivial angular momentum $J$. Remarkably, the angular momentum is not quantized in contrast to the topological case. Furthermore, the angular momentum $J$, as a function of the charge $Q$, has an extreme value and can vanish for certain $Q\neq 0$ (if $n\neq 0$).

During the numerical analysis of the integral characteristics, it was found that solitons under consideration are kinematically stable, even for $n\neq 0$. The stabilization of the solution is provided by the Chern--Simons term since a corresponding coupling constant acts as a mass of the gauge field. For the same reason, the energy of the solution is finite, which ensures localization of the field configuration.

The change of frequency $\omega$ shows the presence of two bounds. At the $\omega\to \omega_\text{min}\neq 0$ limit, the system goes into a thin-wall regime characterized by the large values of energy and charge. The considered charge values indicate the model may be of interest in the context of Bose--Einstein condensation of dipolar quantum gasses \cite{Lahaye:2009qqr}.
For large values of the charge ($Q\gg Q_n$) one can expect the change of $E=E(Q)$ dependence, as it was shown in
\cite{Deshaies-Jacques:2006clf} for $n=0$.

The opposite limit $\omega\to m$ is notable in terms of conformal symmetry restoration \cite{deKok:2008ge}. In this limit, the integral characteristic $E=E(Q)$ ends at an isolated point. Such a picture is typical for the non-relativistic effective scalar field theories. Thus, the particular interest is focused on the possible features related to the presence of the gauge field.

\section*{Acknowledgments}

Numerical studies of gauged non-topological solitons were supported by the grant RSF 22-12-00215-$\Pi$. The work of E. Kim was supported by the Foundation for the Advancement of Theoretical Physics and Mathematics BASIS.

\appendix

\section{Asymptotics of the solitonic solution}\label{app_asymp}

The equations of motion (\ref{f_eq}, \ref{Ga_eq}) are non-linear, and therefore it seems impossible to find their general solution analytically. Nevertheless, one can easily linearize the equations and obtain the following asymptotics for the functions $f$, $G$, $a$ at $r\to\infty$ \cite{Paul:1986ix} (see also \cite{Inozemtsev:1987wv}):
\begin{equation*}
    \big|f(r)-f(\infty)\big|\propto e^{-\sqrt{m^2-\omega^2}\,r},\qquad\big|G(r)-G(\infty)\big|\propto e^{-|g|r},\qquad\big|a(r)-a(\infty)\big|\propto e^{-|g|r}. 
\end{equation*}
Besides, it follows that a relation $|\omega|< m$ should be satisfied, as otherwise the amplitude of the scalar field oscillates at spatial infinity, and localization of the field configuration does not take place.

\begin{samepage}
By representing the ansatz functions in the form of power series about a point $r=0$, one can also find that \cite{Paul:1986ix}
\begin{equation*}
    \big|f(r)-f(0)\big|\propto r^{|n|}+O\big(r^{|n|+2}\big),\qquad \big|G(r)-G(0)\big|\propto r^2+O\big(r^4\big),\qquad \big|a(r)-a(0)\big|\propto r^2+O\big(r^4\big). 
\end{equation*}
The asymptotic expansion for $f$ at $r\to 0$ is necessary for our numerical procedure.
\end{samepage}

\bibliography{biblio}

\end{document}